\documentclass[12pt]{article}
\def\bbbone{{\mathchoice {\rm 1\mskip-4mu l} {\rm 1\mskip-4mu l}
{\rm 1\mskip-4.5mu l} {\rm 1\mskip-5mu l}}}
\def\vP{\vec{P}\, {}^2}
\begin{document}

\title{Doubly Special Relativity and quantum gravity phenomenology}

\author{J. Kowalski-Glikman\thanks{{ Research  partially supported
by the    KBN grant 5PO3B05620.}}\\%
Institute for Theoretical Physics,\\ University of Wroc\l{}aw,\\ Pl. Maxa Borna 9,
50-204 Wroc\l{}aw, Poland \\E-mail: jurekk@ift.uni.wroc.pl}



\maketitle

\begin{abstract}
I review the conceptual, algebraical, and geometrical structure of Doubly Special Relativity. I also
speculate about the possible relevance of DSR for quantum gravity phenomenology.%
\end{abstract}

\section{Introducing DSR}

Without doubts the most challenging problem of high energy physics is to
complete the unfinished revolution that started with the formulation of general
relativity and quantum mechanics in the first decades of twenties century. For
many years however quantum gravity has been associated only with strong
gravitational field regime in the vicinity of space-time singularities, e.g.,
with effects taking place just after (or before) the big bang or inside black
holes. This state of affairs made it impossible to have any reasonable hope to
test quantum gravitational effects in a foreseeable future. It was estimated
that a particle accelerator in which we could test Planck scale scattering
should be of the size of Earth orbit.

The status of quantum gravity phenomenology changed substantially with the
observation\cite{Amelino-Camelia:1999zc}, \cite{Amelino-Camelia:2002vw} that
there is a hope that one may observe imprints of quantum gravity in experiments
that already are or soon will be within the reach of current observational
technologies. Most notably one expects that we may see traces of quantum
gravity in kinematics of ultra-high energy cosmic rays scaterrings (violation
of the GZK bound), as well as in the time-of-flight experiments.

It should be stressed that this new quantum gravity phenomenology paradigm concerns minute
effects of quantum gravity that are present already in the flat space limit.
Moreover these effects are of the kinematical and not dynamical origin, namely they concern
interaction of test particles with flat space-time solution of quantum gravity,
which is assumed to lead to a theory that is somehow different from the
standard Special Relativity.

This observation has led Amelino-Camelia to postulate that there is a new
setting that governs particle kinematics in the flat space-time, which is to
replace Special Relativity. According to \cite{Amelino-Camelia:2000ge},
\cite{Amelino-Camelia:2000mn} such a theory, dubbed ``Doubly (or Deformed) Special
Relativity'' (DSR) should satisfy the following postulates

\begin{itemize}
\item The relativity principle, i.e., equivalence of all inertial observers in
the sense of Galilean Relativity and Special Relativity.

\item The presence of {\em two} observer independent scales: one of velocity $c$,
identified with the speed of light, and second of dimension of mass $\kappa$
(or length), identified with the Planck mass.
\end{itemize}

It should be noted that as an immediate consequence of these postulates 
 DSR theory should possess (like Galilean and Special Relativity theories) a ten
dimensional group of symmetries, corresponding to rotations, boosts, and
translations, which however cannot be just a linear Poincar\'e group as a
result of the presence of the second scale. This immediately poses a problem. Namely,
if we have a theory with observer independent scale of mass, then it follows
that we should expect that the standard Special Relativistic Casimir $E^2 - p^2
= m^2$ is replaced by some nonlinear mass-shell relation between energy and
three-momentum (which would involve the parameter $\kappa$.) But then it
follows that the speed of massless particles would be dependent on the energy
they carry, which makes it hard to understand what would be the operational
meaning of the observer-independent speed of light. In what follows I will
suggest  ways out of this dilemma.

Soon after pioneering papers of Amelino-Camelia it was realized in
\cite{Kowalski-Glikman:2001gp} and \cite{Bruno:2001mw} that the so called
$\kappa$-Poincar\'e algebra \cite{kappaP}, \cite{kappaM} is a perfect
mathematical setting to describe one particle kinematics in DSR. In particular,
in the so-called bicrossproduct basis the commutators between rotations $M_i$,
boosts $N_i$, and the components of momenta $p_\mu$\footnote{As I will argue in
the Section 2 below, it is more precise to think of these quantities as of
Poincar\'e charges carried by the particle.} read
$$
[M_i, M_j] = i\, \epsilon_{ijk} M_k, \quad [M_i, N_j] = i\, \epsilon_{ijk} N_k,
$$
\begin{equation}\label{1}
  [N_i, N_j] = -i\, \epsilon_{ijk} M_k,
\end{equation}
\begin{equation}\label{2}
  [M_i, P_j] = i\, \epsilon_{ijk} P_k, \quad [M_i, P_0] =0,
\end{equation}
\begin{equation}\label{3}
   \left[N_{i}, P_{j}\right] = i\,  \delta_{ij}
 \left( {\kappa\over 2} \left(
 1 -e^{-2{P_{0}/ \kappa}}
\right) + {1\over 2\kappa} \vec{P}\,{}^{ 2}\, \right) - i\, {1\over \kappa}
P_{i}P_{j}
\end{equation}
\begin{equation}\label{4}
\left[N_{i}, P_{0}\right] = i\, P_i
\end{equation}
It is important to note that the algebra  of $M_i$ $N_i$ is
just the standard Lorentz algebra, so one of the first conclusions is that the
Lorentz sector of $\kappa$-Poincar\'e algebra is not deformed. Therefore in DSR
theories, in accordance with the first postulate above, the Lorentz symmetry
{\em is not broken} but merely nonlinearly realized in its action on momenta.
This simple fact has lead some authors (see e.g., \cite{lunoDSR},
\cite{Ahluwalia:2002wf}) to
 claim that DSR is nothing but the standard Special Relativity in non-linear
disguise. As we will see this view is clearly wrong, simply because the algebra
(\ref{1})--({\ref{4}) describes only half of the phase space of the particle,
and the full phase space algebra cannot be reduced to the one of Special
Relativity.

As one can easily check, the Casimir of the $\kappa$-Poincar\'e algebra reads
\begin{equation}\label{5}
\kappa^2\,  \cosh\, \frac{P_0}{\kappa} - \frac{\vec{P}{}^2}2\, e^{P_0/\kappa} = M^2
\end{equation}
from which it follows that the value of three-momentum $|\vec{P}|=\kappa$
corresponds to  infinite energy $P_0=\infty$. One can check easily that in
this particular realization of DSR $\kappa$ is indeed observer independent
\cite{Kowalski-Glikman:2001gp}, \cite{Bruno:2001mw} (i.e., if a particle has momentum 
$|\vec{P}|=\kappa$ for some observer, it has the same momentum for all, Lorentz related, 
observers.) One also sees that the
speed of massless particles, naively defined as derivative of energy over
momentum, increases monotonically with momentum and diverges for the maximal
momentum $|\vec{P}|=\kappa$.

One should note at this point that the bicrossproduct algebra above is not the
only possible realization of DSR. For example, Magueijo and Smolin proposed and
carefully analyzed another DSR proposal in \cite{Magueijo:2001cr},
\cite{Magueijo:2002am}. Moreover there is a basis of DSR, closely related to
the famous Snyder theory \cite{snyder}, in which the energy-momentum space
algebra is purely classical.

\section{Space-time of DSR}

The formulation of DSR in the energy-momentum space is clearly incomplete, as it
lacks any description of the structure of space-time.  DSR
changes somehow our usual perspective that in constructing physical theories we
should start with the ``position space'' picture and only then try to built
the phase space. Here  the starting point is the energy-momentum space, but still for complete understanding of a physical theory
one should know the whole phase space.

There are in principle many ways how the phase space can be constructed. For
example in \cite{Kimberly:2003hp} one constructs the position space along the
same lines as the energy-momentum space has been constructed in
\cite{Magueijo:2001cr}, \cite{Magueijo:2002am}. Here we take another route. We
observe that one of the distinctive features of the $\kappa$-Poincar\'e algebra
is that it possesses additional structures that make it a Hopf algebra. Namely
one can construct the so called co-products for the rotation, boosts, and momentum
generators, which in turn can be use to provide a procedure to construct the
phase space in a unique way.

The co-product is the mapping from an algebra ${\cal A}$ to the tensor product ${\cal A}\otimes {\cal A}$
satisfying some requirements that make it in a sense dual to algebra multiplication,
and for the $\kappa$-Poincar\'e algebra it reads
\begin{equation}\label{6}
 \Delta (P_0) = \bbbone\otimes P_0 + P_0 \otimes \bbbone
\end{equation}
\begin{equation}\label{7}
\Delta (P_k) = P_k\otimes {\rm e}^{-P_0/\kappa} + \bbbone \otimes P_k
\end{equation}
\begin{equation}\label{8}
  \Delta
(M_i) = M_i\otimes \bbbone + \bbbone \otimes M_i
\end{equation}
\begin{equation}\label{9}
  \Delta
(N_i) = \bbbone\otimes N_i + N_i \otimes {\rm e}^{-P_0/\kappa} -
\frac{1}{\kappa}\epsilon_{ijk}\, M_j \otimes P_k
\end{equation}
In order to construct the one-particle phase space we must first introduce
objects that are dual to $M_i$, $N_i$, and $P_\mu$. These are the matrix
$\Lambda^{\mu\nu}$ and the vector $X^\mu$. Let us briefly interpret their
physical meaning. $X^\mu$ are to be dual to momenta $P_\mu$, which clearly
indicates that they should be interpreted as positions, in other words the
generators of translation of momenta. The duality between $\Lambda^{\mu\nu}$
and $M_{\mu\nu} = (M_i, N_i)$ is a bit more tricky. However if one interprets
$M_{\mu\nu}$ in analogy to the interpretation of momenta, i.e., as Lorentz
charge carried by the particle, that is their angular momentum, then the dual
object $\Lambda^{\mu\nu}$ has clear interpretation of Lorentz transformation.
Thus we have the structure of the form $G \times {\cal MP}$, where $G$ is the
Poincar\'e group acting on the space of Poincar\'e charges of the particle
${\cal MP}$. We see therefore that we can make use with the powerful
mathematical theory of Lie-Poisson groups and co-adjoint orbits (see, for
example, \cite{kirillov:1976}, \cite{alekseev:1994}) and their quantum deformations.

Following \cite{kosinski1995} and \cite{luno} we assume the following form of
the co-product on the group
\begin{equation}\label{10}
 \Delta(X^{\mu})=\Lambda^{\mu}{}_{\nu}\otimes
X^{\nu}+X^{\mu}\otimes \bbbone
\end{equation}
and
\begin{equation}\label{11}
\Delta(\Lambda^{\mu}{}_{\nu})=\Lambda^{\mu}{}_{\rho}\otimes
\Lambda^{\rho}{}_{\nu}
\end{equation}
The next step is to define the pairing between elements of the algebra and the
ones of the group in a canonical way (note however that in 3 space-time
dimensions another canonical pairing is possible, see below) that establish 
the duality between these two structures.
\begin{equation}\label{12}
 <P_{\mu},X^{\nu}>= i\delta_{\mu}^{\nu}
\end{equation}
\begin{equation}\label{13}
  <\Lambda^{\mu}{}_{\nu},M^{\alpha\beta}>= i
\left(g^{\alpha\mu}\delta^{\beta}_{\nu}-g^{\beta\mu}\delta^{\alpha}_{\nu}\right)
\end{equation}
\begin{equation}\label{14}
  <\Lambda^{\mu}{}_{\nu},1>= \delta^{\mu}_{\nu}
\end{equation}
In (\ref{13}) $g^{\alpha\mu}$ is the Minkowski space-time metric. This pairing
must be consistent with the co-product structure in the following sense
\begin{equation}\label{16a}
  <A, XY> = <A_{(1)}, X><A_{(2)}, Y>,
\end{equation}
 \begin{equation}\label{16b}
 <AB,X> =<A, X_{(1)}><B, X_{(2)}>,
\end{equation}

The rules (\ref{12})--(\ref{16b}) make it possible to construct the commutator
algebra of the phase space. To this end one makes use of the Heisenberg double
procedure \cite{alekseev:1994}, \cite{luno}, that defines the brackets in terms
of the pairings as follows
\begin{equation}\label{15}
 \left[X^{\mu},P_{\nu}\right]= P_{\nu(1)}\left< X^{\mu}_{(1)},P_{\nu(2)}\right> X^{\mu}_{(2)}
-P_{\nu}X^{\mu},
\end{equation}
\begin{equation}\label{16}
  \left[X^\mu,M^\rho{}_\sigma\right]= M_{(1)}{}^\rho{}_\sigma \left< X_{(1)}{}^\mu,M_{(2)}{}^\rho{}_\sigma \right> X_{(2)}{}^\mu - M^\rho{}_\sigma X^\mu,
\end{equation}
and analogously for $\Lambda^{\mu}{}_{\nu}$ commutators, where on the right
hand side we make use of the standard (``Sweedler'') notation for co-product
$$
\Delta {\cal T} = \sum {\cal T}_{(1)} \otimes {\cal T}_{(2)}.
$$

As an example let us perform these steps in the bicrossproduct DSR. It follows
from (\ref{7}), and (\ref{12}), and (\ref{16b}) that
$$
<P_i, X_0 X_j> = -\frac1\kappa\, \delta_{ij}, \quad <P_i,  X_jX_0>=0,
$$
from which one gets
\begin{equation}\label{17a}
[X_0, X_i] = -\frac{i}\kappa\, X_i.
\end{equation}
Similarly, using (\ref{15}) we get the standard relations
\begin{equation}\label{17b}
[P_0, X_0] = -i, \quad [P_i, X_j] = i \, \delta_{ij}.
\end{equation}
It turns out that the phase space algebra contains one more non-vanishing
commutator (which can be, of course, also obtained from Jacobi identity),
namely
\begin{equation}\label{17c}
 [P_i, X_0] = -\frac{i}\kappa\, P_i.
\end{equation}

Thus we have constructed the phase space of the bicrossproduct DSR. Let us
stress that this construction relies heavily on the form of co-product.
However, as it will turn out in the next section, some of the commutators are
sensitive to the particular form of the DSR, while the others are not. In
particular we will see that the non-commutativity of positions (\ref{17a}) is
to large extend universal for a whole class of DSR theories. The
non-commutative space-time with such Lie-like type of non-commutativity is
called $\kappa$-Minkowski space-time.

\section{From DSR theory to DSR theories}

The introduction of invariant momentum (or mass) scale $\kappa$ has immediate
consequences.  The most important is that there is nothing sacred about the
bicrossproduct DSR presented in the last two sections, as one can simply use
$\kappa$ to define new energy end momentum as analytic functions of the old
ones, to wit
\begin{equation}\label{18}
{ \cal P}_i = {\cal F}_i(P_i, P_0; \kappa), \quad { \cal P}_0 = {\cal F}_0(P_i,
P_0; \kappa).
\end{equation}
 Observe that such a possibility is excluded in a theory without
any mass scale, like special relativity and Newtonian mechanics,  in which the
energy momentum space is linear, and the mass shell condition is expressed by
quadratic form. Therefore there are two possibilities to to answer the natural
question: which momenta are the ``right'' ones? Namely one may hope that the
theory of quantum gravity or some other fundamental theory, from which DSR is
descending will tell what is the right choice. Second one can contemplate the
possibility that in the final, complete formulation of DSR one will have to do
with some kind of ``energy-momentum general covariance'', i.e., that physical
observables do not depend on a particular realization of eq.~(\ref{18}), like
observables in general relativity do not depend on coordinate system. Then a
natural question arises: is it possible to understand transformations
(\ref{18}) as coordinate transformations on some (energy-momentum) space?

Surprisingly enough the answer to this question is in positive: indeed the
transformations between DSR theories, described by (\ref{18}) are nothing but
coordinate transformation of the constant curvature manifold, on which momenta
live. To reach this conclusion one observes first
\cite{Kowalski-Glikman:2002we}, \cite{Kowalski-Glikman:2002jr} that it follows
from the Heisenberg double construction that both the $\kappa$-Minkowski
commutator (\ref{17a}) and the commutators between Lorentz charges $M_{\mu\nu}$
and positions $X_\mu$ are left invariant by the transformations (\ref{18}).
This follows from the fact that the transformations (\ref{18}) a severely
constrained by assumed rotational invariance and the fact that in the $\kappa\rightarrow\infty$ limit we must get the standard Special Relativity, and in the leading order read
\begin{equation}\label{19}
{ \cal P}_i \approx P_i + \alpha\, \frac1\kappa\, P_i P_0 +
O\left(\frac1{\kappa^2}\right), \quad { \cal P}_0 = P_0 + \beta\,
\frac1\kappa\, P_0^2 + O\left(\frac1{\kappa^2}\right)
\end{equation}
where $\alpha$ and $\beta$ are numerical parameters. It turns out that in
computing   $X$ and $M$ commutators Heisenberg double procedure picks up only
the first terms in this expansion, and thus the form of the commutators remains
unchanged. Of course, the position-momenta commutators are changed by
(\ref{18}), (\ref{19}).

Next it was realized in \cite{Kowalski-Glikman:2002ft},
\cite{Kowalski-Glikman:2003we} that the algebra of positions and Lorentz
charges is nothing but the de Sitter SO(4,1) algebra. It follows that one can
identify the energy--momentum space with the four dimensional de Sitter space.
This space can be constructed as a four dimensional  surface of constant
curvature $\kappa$ in the five dimensional Minkowski space with coordinates
$\eta_A$, $A=0,\ldots,4$, to wit
\begin{equation}\label{20}
-\eta_0^2 + \eta_1^2 + \cdots + \eta_4^2 = \kappa^2.
\end{equation}
Positions $X_\mu$ and Lorentz charges $M_{\mu\nu}$ act on $\eta_A$ variables as
follows
\begin{equation}\label{21}
  [X_0,\eta_4] = \frac{i}\kappa\, \eta_0, \quad [X_0,\eta_0] = \frac{i}\kappa\, \eta_4, \quad [X_0,\eta_i] = 0,
\end{equation}
\begin{equation}\label{22}
  [X_i, \eta_4] = [X_i, \eta_0] =\frac{i}\kappa\, \eta_i, \quad [X_i, \eta_j] = \frac{i}\kappa\,
\delta_{ij}(\eta_0 - \eta_4),
\end{equation}
and
\begin{equation}\label{23}
  [M_i, \eta_j] = i\epsilon_{ijk}\eta_k, \quad [N_i, \eta_j] = i\, \delta_{ij}\, \eta_0, \quad [N_i, \eta_0] = i\,  \eta_i,
\end{equation}
It should be noted that there is another decomposition of $SO(4,1)$ generators
\cite{Kowalski-Glikman:2002ft}, \cite{Kowalski-Glikman:2003we}, in which the
resulting algebra is exactly the one considered by Snyder \cite{snyder}.

On the space (\ref{20}) one can built various co-ordinate systems, each related
to some DSR theory. In particular, one recovers the bicrossproduct DSR with
the following coordinates (which are, accidentally, the standard ``cosmological'' coordinates
on de Sitter space)
\begin{eqnarray}
{\eta_0} &=& \kappa\, \sinh \frac{P_0}\kappa + \frac{\vec{P}\,{}^2}{2\kappa}\,
e^{  \frac{P_0}\kappa} \nonumber\\
\eta_i &=&   P_i \, e^{  \frac{P_0}\kappa} \nonumber\\
{\eta_4} &=&  \kappa\, \cosh \frac{P_0}\kappa  - \frac{\vec{P}\,{}^2}{2\kappa}
\, e^{  \frac{P_0}\kappa}.   \label{24}
\end{eqnarray}
Using (\ref{24}), (\ref{22}), and the Leibnitz rule, one easily recovers the
commutators (\ref{1})--(\ref{4}).

Other coordinates systems, are possible, of course. In particular one may
choose the ``standard basis'' in which
\begin{equation}\label{25}
 {\cal P}_\mu = \eta_\mu.
\end{equation}
Note that in this basis (or classical DSR) the commutators of all Poincar\'e
charges, ${\cal P}_\mu$ and $M_{\mu\nu}$ are purely classical. However, the
positions commutators are still non-trivial, as well as the momenta/positions
cross-relations. This means that the (observer-independent) scale $\kappa$
disappears completely from the Lorentz sector, but is still present in the
translational one. Thus such a theory fully deserves the name DSR. As we will
see classical DSR diverges from Special Relativity radically in the many
particles sector.

The de Sitter space setting reveals the geometrical structure of DSR theories. As
we saw the energy momentum  space of DSR is a four dimensional manifold of
positive constant curvature, and the curvature radius equals the scale
$\kappa$. The Lorentz charges and positions are identified with the set of ten
tangent vectors to the de Sitter energy-momentum space, and as an immediate
consequence of this their algebra is independent of any particular coordinate
system on this space.  However the latter seems to be, at least naively,
physically relevant. Each such coordinate system defines for us (up to the
redundancy discussed in \cite{Kowalski-Glikman:2003we}) the physical  energy and momentum. In one-particle sector the particular choice may not be
relevant, but it seems that it would be of central importance for the proper
understanding of many particles phase spaces, in particular in analysis of the
phenomenologically important issue of particles scattering and conservation
laws.

Having obtained the one-particle phase space of DSR, it is natural to proceed
with construction of the field theory. Here two approaches are possible. One
can try to construct field theory on the non-commutative $\kappa$-Minkowski
space-time. Attempts to construct such a theory has been reported, for example,
in \cite{Kosinski:1999ix}, \cite{Kosinski:1999dw}, \cite{Kosinski:2001ii},
\cite{Kosinski:2003xx}, \cite{Amelino-Camelia:2001fd},
\cite{Amelino-Camelia:2002mu}. This line of research is, however, far from
being able to give any definite results, though some partial results, like an
interesting, nontrivial vertex structure reported in
\cite{Amelino-Camelia:2001fd}, \cite{Amelino-Camelia:2002mu} may shed some
light on physics of the scattering processes. The major obstacle seems to be
lack of the understanding of functional analysis on the spaces with Lie-type of
non-commutativity, which is most likely a deep and hard mathematical problem
(already the definition of appropriate differential and integral calculi is a
mater of discussion.) Therefore it seems simpler (and in fact more along the
line of the DSR proposal, where the energy momentum space is more fundamental
than the space-time structures) to try to built (quantum) field theory in
energy-momentum space directly. This would amount to understand how to define
(quantum) fields on the curved energy-momentum space, but, in principle, for
spaces of constant curvature at least functional analysis is well understood.
It should be noted that such an idea has been contemplated for a long time, and
in fact it was one of the main motivations of \cite{snyder}. Field theories
with curved energy-momentum manifold has been intensively investigated by Kadyshevsky 
and others \cite{Kadyshevsky}, without any conclusive results, though.

Finally it should be noted that one can construct DSR theories with
energy-momentum space being the four dimensional space of constant negative
curvature, i.e, anti de Sitter space, as well as the flat space. In this latter case,
the positions still form $\kappa$-Minkowski space-time, but one can construct four commuting
pseudo-position variables, as linear combinations of $\kappa$-Minkowski positions and Lorentz generators.
This results, as well as more detailed
investigations of phase spaces of DSR are reported in the recent paper \cite{Blaut:2003wg}.

\section{One-particle quantum gravity phenomenology: time-of-flight experiment}

One of the simplest experimental tests of quantum gravity phenomenology is the time-of-flight experiment. In this experiment which is to be performed in a near future with good accuracy by the GLAST satellite (see e.g.,
\cite{Amelino-Camelia:2002vw} and references therein) one measures the energy-dependence of velocity of light coming from a distant source. Naively, most DSR models predicts positive signal in such an experiment. Indeed, in DSR $\partial E(p)/\partial p$ is not constant, which suggest that velocity of massless particles may depend on the energy they carry. This is the case, for example, both in the bicrossproduct DSR and in the Magueijo-Smolin model.

In the more careful analysis reported in \cite{Amelino-Camelia:2002tc} the authors construct the wave packet from plane waves moving on the $\kappa$-Minkowski space-time, and then calculate the group velocity of such a packet, which turns-out to be exactly $v^{(g)} = \partial E(p)/\partial p$. This result is puzzling in view of the phase space calculation of velocity, which I will present below, and discuss  later in this section.

Let us try to compute velocity starting from the phase space of DSR theories. This computation has been presented in \cite{Daszkiewicz:2003yr} (see also \cite{Kosinski:2002gu} and \cite{Mignemi:2003ab}).

The idea is to start with the commutators (\ref{21})--(\ref{23}). Note first that since  
the for the variable $\eta_4$, $[M_i, \eta_4]=[N_i,\eta_4]=0$,  $\kappa\, \eta_4$ is a Casimir 
(cf.~(\ref{24}) and can be therefore naturally identified with the relativistic Hamiltonian for
free particle in any DSR basis. Indeed it
is by construction Lorentz-invariant, and reduces to the standard relativistic
particle hamiltonian in the large $\kappa$ limit. Moreover, using the fact that
for $P_\mu$ small compared to $\kappa$, in any DSR theory $\eta_\mu \sim P_\mu
+ O(1/\kappa)$ we have
\begin{equation}\label{a}
 \kappa\eta_4 = \kappa^2\sqrt{1 + \frac{P_0^2 -\vP}{\kappa^2}} \sim \kappa^2 + \frac12\left(P_0^2 -\vP\right) +
 O\left(\frac1{\kappa^2}\right)
\end{equation}

Then it follows from eq.~(\ref{22}) that $\eta_\mu = [x_\mu,
\kappa\eta_4]$ can be identified with four velocities $u_\mu$. The Lorentz
transformations of four velocities are then given by eq.~(\ref{23}) and are identical
with those of  Special Relativity. Moreover, since
\begin{equation}\label{b}
 u_0^2 - \vec{u}\,{}^2 \equiv {\cal C} = M^2
\end{equation}
by the standard argument the three velocity equals $v_i = u_i/u_0$ and the
speed of massless particle equals $1$.  Let me stress here once again that this
result is DSR model independent, though, of course, the relation between three
velocity of massive particles and energy they carry depends on a particular DSR
model one uses.

Thus this calculation indicates that GLAST should not see any signal of energy 
dependent speed of light, at least if it is correct  
to think of photons as of point massless classical particles, as I have implicitly assumed here.

As I mentioned above there is a puzzling discrepancy between this calculation and the 
one reported in \cite{Amelino-Camelia:2002tc}, where the authors analyzed plane waves propagation in
non-commutative $\kappa$-Minkowski space-time. However one may argue that without well 
quantum understood field theory, which would incorporate DSR somehow, it is hard to understand what role
(if any) plane waves would play in such a theory. In particular it is far from obvious that such plane
waves could be interpreted as one-particle solutions of DSR-field theory, and thus that they are correct
tools to describe photons.

It should be stressed that the issue of velocity of {\em physical} particles is not settled on the theoretical ground, and thus any experimental input would be extremely valuable.

\section{Remarks on multi-particle systems}

Having obtained the one-particle phase space of DSR it is natural to try to
generalize this result to find the two- and multi-particles phase spaces. It
turns out however that such a generalization is very difficult, and in spite of
many attempts not much about multi-particles kinematics is known. On the other
hand the control over particle scattering processes is of utmost relevance in the
analysis of seemingly one of the most important windows to quantum gravity
phenomenology, provided by Ultra High Energy Cosmic Rays and possible
violations of predictions of Special Relativity in UHECR physics (see e.g.,
\cite{Amelino-Camelia:2002vw} for more detailed discussion and the list of
relevant references.)

Ironically,  we have in our disposal the mathematical structure that seem to
provide a tool to solve multi-particle the problem directly. This structure is
co-product. Recall that the co-product is a mapping from the algebra to its
tensor product
\begin{equation}\label{26}
\Delta: {\cal A} \rightarrow {\cal A}\otimes {\cal A}
\end{equation}
 and thus it provides the rule how the algebra acts on tensor products of its
representations. We know that in ordinary quantum mechanics two-particles
states are described as a tensor product of single-particle ones. Note that
this is a very strong physical assumption: in making it we claim that any
two-particle system is nothing but two particles in a black box, i.e., that the
particles preserve their identities even in multi-particle states. But it is
well possible that multi-particle states differ qualitatively from the
single-particle ones, for example as a result of non-local interactions. Let us
however assume that in  DSR to obtain the multi-particle states one should only
tensor the single-particle ones, and let us try to proceed.

In the case of classical groups the co-product is trivial: 
$\Delta{G} = G\otimes1 + 1\otimes G$ which means that the group action on two particle
states just respects Leibnitz rule. For example the total momentum of two
particles in Special relativity is just the sum of their momenta:
$$
\Delta(P_\mu)\, |1+2> = \Delta(P_\mu)\, |P^{(1)}> \otimes\, |P^{(2)}> =$$
\begin{equation}\label{27}
\left(P_\mu\otimes\bbbone + \bbbone\otimes P_\mu\right)\,|P^{(1)}> \otimes\,
|P^{(2)}> =\left(P^{(1)}_\mu + P^{(2)}_\mu\right)\,|P^{(1)}> \otimes\,
|P^{(2)}>
\end{equation}
 In the case
of quantum algebras the co-product is non-trivial by definition (if it was
trivial we would have to do with a classical group). This immediately leads to
the problem, as I will argue below.

Before turning to this let us point out yet another problem, relevant for the
bicrossproduct DSR. Namely the co-product has been constructed so that
two-particle states transform as the single-particle ones (for example in
Special Relativity total momentum is Lorentz vector.) But then it follows that
total momentum must satisfy the same mass shell relation as the single particle
does. We know however that in the case of the bicrossproduct DSR we have to do
with maximal momentum for particles, of order of Planck mass. While  acceptable
for Planck scale elementary particles, this is certainly violated for
macroscopic bodies. This nice quantum gravity phenomenology experiment can be
easily performed by everyone just by kicking a soccer ball! So we know that
there is an experimental proof that either our procedure of attributing
momentum to composite system by tensoring and applying co-product, or the
bicrossproduct DSR, or both are wrong.

To investigate things further let us turn to the DSR theory, which does not
suffer from the ``soccer ball problem'' namely to the classical basis DSR with
standard dispersion relation ${\cal P}_0^2 - {\cal P}_i^2 = m^2$, in which the
de Sitter coordinates are given by (\ref{25}). The co-product for this basis has been calculated in
\cite{Kowalski-Glikman:2002jr} and up to the leading terms in $1/\kappa$
expansion read
  \begin{equation}\label{28}
\Delta({\cal P}_{0}) = \bbbone\otimes {\cal P}_{0} + {\cal P}_{0}\otimes
\bbbone + \frac1{\kappa}\, {\cal P}_i \otimes {\cal P}_i + \ldots
\end{equation}
\begin{equation}\label{29}
 \Delta({\cal P}_{i}) =\bbbone\otimes {\cal P}_{i} + {\cal P}_{i}\otimes
\bbbone + \frac1{\kappa}\, {\cal P}_0 \otimes {\cal P}_i + \ldots
\end{equation}
Using this we see that the total momentum of two-particles system
\begin{equation}\label{30}
{\cal P}_{0}^{(1+2)} = {\cal P}_{0}^{(1)} + {\cal P}_{0}^{(2)} +
\frac1{\kappa}\,{\cal P}_{i}^{(1)}  {\cal P}_{i}^{(2)}
\end{equation}
\begin{equation}\label{31}
{\cal P}_{i}^{(1+2)} = {\cal P}_{i}^{(1)} + {\cal P}_{i}^{(2)} +
\frac1{\kappa}\,{\cal P}_{0}^{(1)}  {\cal P}_{i}^{(2)}
\end{equation}
As it stands, the formulas (\ref{30}, \ref{31}) suffer from two problems: first
of all, recalling that ${\cal P}_{\mu}$ transforms as a Lorentz vector for
single particle, these expressions look terribly non-covariant. Second, even
though (\ref{30}) is symmetric in exchanging particles labels $1
\leftrightarrow 2$, (\ref{31}) is not. How do we know which particle is first
and which is second? Let us try to resolve these puzzles in turn.

That the first puzzle is just an apparent paradox follows immediately from the
consistency of the quantum algebra. As I said above the action of boosts on
two-particle state is such that total momentum transforms exactly as the
single-particle momentum does. This is in fact the very reason of the ``soccer
ball problem'' in the bicrossproduct DSR. In fact the boosts do not only act on
${\cal P}_{\mu}^{(1)}$ and ${\cal P}_{\mu}^{(2)}$ independently; they also mix
them in a special way. To see this, note that boosts also act on two-particle states
by co-product, therefore in order to find out how a two-particle state changes
when we boost it we must compute the commutator $[\Delta(N), \Delta({\cal
P})]$. Recall now that the co-product of boosts reads (again up to the leading
terms in $1/\kappa$ expansion)
\begin{equation}\label{32}
  \Delta
(N_i) = \bbbone\otimes N_i + N_i \otimes \bbbone - \frac{1}{\kappa}\, N_i
\otimes {\cal P}_0 - \frac{1}{\kappa}\, \epsilon_{ijk}\, M_j \otimes {\cal P}_k
\end{equation}
Using this one easily checks explicitly that
\begin{equation}\label{33}
[\Delta(N_i), \Delta({\cal P}_j)] = \delta_{ij} \Delta({\cal P}_0), \quad
[\Delta(N_i), \Delta({\cal P}_0)] = \Delta({\cal P}_i)
\end{equation}
from which it follows that ${\cal P}_{0}^{(1+2)}$ and ${\cal P}_{i}^{(1+2)}$ do
transform covariantly, as they should. Of course equation (\ref{33}) holds to
all orders, as it just reflects the defining property of the co-product.

Let us now turn to the second puzzle. Here I have much less to say, as this paradox has 
not been yet solved. One should however mention an interesting result obtained in the case
of the analogous problem in deformed, non-relativistic model. In the paper \cite{kosinski:1998} 
the authors find that even though there is an apparent asymmetry in particle labels due to the
asymmetry of the co-product, the representations with flipped labels are related to the original 
ones by unitary transformation, and are therefore completely equivalent. In the similar spirit
in \cite{Bonechi:cb} one uses the fact of such an equivalence in 1+1 dimensions to demand that 
the action of generators on two particles (bosonic) states is through symmetrized co-product.

\section{2+1 (quantum) gravity as an example of DSR theory}

As I stressed above DSR theory has a number of conceptual problems. It is therefore quite important to find a model sharing features with DSR, which may shed some light on these conceptual problems. Surprisingly enough such a toy model exists, namely the 2+1 dimensional gravity coupled to point particles is nothing but a particular realization of DSR in 2+1 dimensions.

The relation of DSR theory in 2+1 dimensions and the 2+1 gravity has been pointed out in \cite{Amelino-Camelia:2003xp} and \cite{Freidel:2003sp}. In particular it has been shown in \cite{Freidel:2003sp} that the phase space of single particle in DSR is isomorphic to the phase space of a particle in 2+1 dimensional gravity \cite{Matschull:1997du}. Thus for a single particle the three dimensional DSR and three dimensional gravity coupled to the particle are just identical theories. One can extend this to multi-particle sector, assuming that also in this case the 2+1 gravity coupled to particles {\em is} a particular realization of the DSR. Then one can translate results obtained in the framework of 2+1 gravity, as  the analysis of particle scattering \cite{Louko:2001ed}, and the natural emergence of quantum groups (see, e.g.\ \cite{Bais:2002ye}, \cite{Meusburger:2003ta} and references therein) to the DSR language. This is extremely important because 2+1 gravity setting with well understood operational definitions of concepts  makes it possible to address many conceptual issues of DSR.

Last but not least the lesson from 2+1 gravity makes it possible to claim that in 3+1 dimensions DSR is a descendant of quantum gravity. Namely 3+1 dimensional quantum gravity should certainly admit in some limit solutions corresponding to flat space with point particles moving on it. Given the lesson from 2+1 dimensions it is not unreasonable to expect that in such a limit theory of gravity would become a topological field theory, whose solution will be a locally flat space with punctures corresponding to particles, whose gravitational interactions would reduce to a kine of topological interactions, as in 2+1 gravity. The claim is that the effective theory governing particle kinematics in this regime, and thus a physical theory of particle kinematics in the (trans)-Planckian regime, will be some form of DSR.

\section*{Acknowledgment} I would like to thank my friends and
collaborators  Giovanni Amelino-Camelia, Arkadiusz B\l{}aut, Marcin
Daszkiewicz, Laurent Freidel, Katarzyna Imi\l{}kowska, Jerzy Lukierski, Joao Magueijo, Sebastian Nowak,
Artem Starodubtsev, and Lee Smolin for their patience in explaining new ideas, pointing out my
mistakes, and numerous discussions.


\begin{thebibliography}{99}
\bibitem{Amelino-Camelia:1999zc}
G.~Amelino-Camelia, ``Are we at the dawn of quantum-gravity phenomenology?,''
Lect.\ Notes Phys.\  {\bf 541} (2000) 1 [arXiv:gr-qc/9910089].

\bibitem{Amelino-Camelia:2002vw}
G.~Amelino-Camelia, ``Quantum-gravity phenomenology: Status and prospects,''
Mod.\ Phys.\ Lett.\ A {\bf 17} (2002) 899 [arXiv:gr-qc/0204051].

\bibitem{Amelino-Camelia:2000ge}
G.~Amelino-Camelia, ``Testable scenario for relativity with minimum-length,''
Phys.\ Lett.\ B {\bf 510} (2001) 255 [arXiv:hep-th/0012238].

\bibitem{Amelino-Camelia:2000mn}
G.~Amelino-Camelia, ``Relativity in space-times with short-distance structure
governed by an observer-independent (Planckian) length scale,'' Int.\ J.\ Mod.\
Phys.\ D {\bf 11} (2002) 35 [arXiv:gr-qc/0012051].

\bibitem{Kowalski-Glikman:2001gp}
J.~Kowalski-Glikman, ``Observer independent quantum of mass,'' Phys.\ Lett.\ A
{\bf 286} (2001) 391 [arXiv:hep-th/0102098].



\bibitem{Bruno:2001mw}
N.~R.~Bruno, G.~Amelino-Camelia and J.~Kowalski-Glikman,
``Deformed boost
transformations that saturate at the Planck scale,'' Phys.\ Lett.\ B {\bf 522}
(2001) 133 [arXiv:hep-th/0107039].

\bibitem{kappaP} J.~Lukierski, H.~Ruegg, A.~Nowicki and V.~N.~Tolstoi,
``Q deformation of Poincare algebra,'' Phys.\ Lett.\ B {\bf 264} (1991) 331.
\bibitem{kappaM} S.~Majid and H.~Ruegg, ``Bicrossproduct structure of kappa Poincare group
and noncommutative geometry,'' Phys.\ Lett.\ B {\bf 334} (1994) 348
[arXiv:hep-th/9405107]; J.~Lukierski, H.~Ruegg and W.~J.~Zakrzewski,
``Classical and quantum mechanics of free kappa relativistic systems,'' Annals
Phys.\  {\bf 243} (1995) 90 [arXiv:hep-th/9312153].

\bibitem{lunoDSR} J.~Lukierski and A.~Nowicki,
``Doubly Special Relativity versus $\kappa$-deformation of relativistic
kinematics,'' Int.\ J.\ Mod.\ Phys.\ A {\bf 18} (2003) 7
[arXiv:hep-th/0203065].

\bibitem{Ahluwalia:2002wf}
D.~V.~Ahluwalia--Khalilova, ``Operational indistinguishabilty of doubly special
relativities from  special relativity,'' arXiv:gr-qc/0212128.

\bibitem{Magueijo:2001cr}
J.~Magueijo and L.~Smolin, ``Lorentz invariance with an invariant energy
scale,'' Phys.\ Rev.\ Lett.\  {\bf 88} (2002) 190403 [arXiv:hep-th/0112090].

\bibitem{Magueijo:2002am}
J.~Magueijo and L.~Smolin, ``Generalized Lorentz invariance with an invariant
energy scale,'' Phys.\ Rev.\ D {\bf 67} (2003) 044017 [arXiv:gr-qc/0207085].

\bibitem{snyder} H.~S.~Snyder,
``Quantized Space-Time,'' Phys.\ Rev.\  {\bf 71} (1947) 38.

\bibitem{Kimberly:2003hp}
D.~Kimberly, J.~Magueijo and J.~Medeiros, ``Non-Linear Relativity in Position
Space,'' arXiv:gr-qc/0303067.

\bibitem{kirillov:1976} A.~A.~Kirillov, {\em Elements of the Theory of
Representations}, Springer 1976.

\bibitem{alekseev:1994} A.~Yu.~Alekseev and A.~Z.~Malkin, ``Symplectic
structures associated with Lie-Poisson groups'', Comm.\ Math.\ Phys.\ {\bf 162}
(1994) 147.

\bibitem{kosinski1995}
P.~Kosinski and P.~Maslanka, ``The $\kappa -$Weyl group and its algebra'',
arXiv:q-alg/9512018.

\bibitem{luno} J. Lukierski and A. Nowicki, `` Heisenberg double description of $\kappa$-Poincar\'e
 algebra and $\kappa$-deformed phase space'', Proceedings of
Quantum Group Symposium at Group 21, (July 1996, Goslar) Eds. H.-D. Doebner and
V.K. Dobrev, Heron Press, Sofia, 1997, p. 186, [arXiv:q-alg/9702003].

\bibitem{Kowalski-Glikman:2002we}
J.~Kowalski-Glikman and S.~Nowak, ``Doubly special relativity theories as
different bases of kappa-Poincare  algebra,'' Phys.\ Lett.\ B {\bf 539} (2002)
126 [arXiv:hep-th/0203040].

\bibitem{Kowalski-Glikman:2002jr}
J.~Kowalski-Glikman and S.~Nowak, ``Non-commutative space-time of doubly
special relativity theories,'' Int.\ J.\ Mod.\ Phys.\ D {\bf 12} (2003) 299
[arXiv:hep-th/0204245].

\bibitem{Kowalski-Glikman:2002ft}
J.~Kowalski-Glikman, ``De Sitter space as an arena for doubly special
relativity,'' Phys.\ Lett.\ B {\bf 547} (2002) 291 [arXiv:hep-th/0207279].

\bibitem{Kowalski-Glikman:2003we}
J.~Kowalski-Glikman and S.~Nowak, ``Doubly special relativity and de Sitter
space,'' Class.\ Quant.\ Grav.\ {\bf 20} (2003) 4799 [arXiv:hep-th/0304101].

\bibitem{Kosinski:1999ix}
P.~Kosinski, J.~Lukierski and P.~Maslanka, ``Local D = 4 field theory on
kappa-deformed Minkowski space,'' Phys.\ Rev.\ D {\bf 62} (2000) 025004
[arXiv:hep-th/9902037].





\bibitem{Kosinski:1999dw}
P.~Kosinski, J.~Lukierski and P.~Maslanka, ``Local field theory on
kappa-Minkowski space, star products and  noncommutative translations,''
Czech.\ J.\ Phys.\  {\bf 50} (2000) 1283 [arXiv:hep-th/0009120].



\bibitem{Kosinski:2001ii}
P.~Kosinski, J.~Lukierski and P.~Maslanka, ``kappa-deformed Wigner construction
of relativistic wave functions and  free fields on kappa-Minkowski space,''
Nucl.\ Phys.\ Proc.\ Suppl.\  {\bf 102} (2001) 161 [arXiv:hep-th/0103127].

\bibitem{Kosinski:2003xx}
P.~Kosinski, P.~Maslanka, J.~Lukierski and A.~Sitarz, ``Generalized
kappa-deformations and deformed relativistic scalar fields on noncommutative
Minkowski space,'' arXiv:hep-th/0307038.



\bibitem{Amelino-Camelia:2001fd}
G.~Amelino-Camelia and M.~Arzano, ``Coproduct and star product in field
theories on Lie-algebra  non-commutative space-times,'' Phys.\ Rev.\ D {\bf 65}
(2002) 084044 [arXiv:hep-th/0105120].

\bibitem{Amelino-Camelia:2002mu}
G.~Amelino-Camelia, M.~Arzano and L.~Doplicher, ``Field theories on canonical
and Lie-algebra noncommutative spacetimes,'' arXiv:hep-th/0205047.

\bibitem{Kadyshevsky} V.~G.~Kadyshevsky et.~al., ``Quantum field theory and a
new universal high-energy scale'', part I, Nuovo Cim.\ {\bf 87 A} (1985) 324;
part II, Nuovo Cim.\ {\bf 87 A} (1985) 350; part III, Nuovo Cim.\ {\bf 87 A}
(1985) 373, and references therein.

\bibitem{Blaut:2003wg}
A.~Blaut, M.~Daszkiewicz, J.~Kowalski-Glikman and S.~Nowak,
``Phase spaces of Doubly Special Relativity,''
arXiv:hep-th/0312045.



\bibitem{Amelino-Camelia:2002tc}
G.~Amelino-Camelia, F.~D'Andrea and G.~Mandanici, ``Group velocity in
noncommutative spacetime,'' arXiv:hep-th/0211022.

\bibitem{Daszkiewicz:2003yr}
M.~Daszkiewicz, K.~Imilkowska and J.~Kowalski-Glikman,
``Velocity of particles in doubly special relativity,''
arXiv:hep-th/0304027.

\bibitem{Kosinski:2002gu}
P.~Kosinski and P.~Maslanka,
``On the definition of velocity in doubly special relativity theories,''
Phys.\ Rev.\ D {\bf 68} (2003) 067702
[arXiv:hep-th/0211057].



\bibitem{Mignemi:2003ab}
S.~Mignemi,
``On the definition of velocity in theories with two observer-independent  scales,''
Phys.\ Lett.\ A {\bf 316} (2003) 173
[arXiv:hep-th/0302065].



\bibitem{kosinski:1998} P.~Kosinski and P.~Maslanka,  ``Deformed Galilei symmetry,'' [arXiv:math.QA/9811142].

\bibitem{Bonechi:cb}
F.~Bonechi, E.~Celeghini, R.~Giachetti, E.~Sorace and M.~Tarlini,
``Quantum Galilei Group As Symmetry Of Magnons,''
Phys.\ Rev.\ B {\bf 46} (1992) 5727
[arXiv:hep-th/9203048].

\bibitem{Amelino-Camelia:2003xp}
G.~Amelino-Camelia, L.~Smolin and A.~Starodubtsev,
``Quantum symmetry, the cosmological constant and Planck scale  phenomenology,''
arXiv:hep-th/0306134.

\bibitem{Freidel:2003sp}
L.~Freidel, J.~Kowalski-Glikman and L.~Smolin,
``2+1 gravity and doubly special relativity,''
arXiv:hep-th/0307085.

\bibitem{Matschull:1997du}
H.~J.~Matschull and M.~Welling,
``Quantum mechanics of a point particle in 2+1 dimensional gravity,''
Class.\ Quant.\ Grav.\  {\bf 15} (1998) 2981
[arXiv:gr-qc/9708054].

\bibitem{Louko:2001ed}
J.~Louko and H.~J.~Matschull,
``The 2+1 Kepler problem and its quantization,''
Class.\ Quant.\ Grav.\  {\bf 18} (2001) 2731
[arXiv:gr-qc/0103085].

\bibitem{Bais:2002ye}
F.~A.~Bais, N.~M.~Muller and B.~J.~Schroers,
``Quantum group symmetry and particle scattering in (2+1)-dimensional  quantum gravity,''
Nucl.\ Phys.\ B {\bf 640} (2002) 3
[arXiv:hep-th/0205021].

\bibitem{Meusburger:2003ta}
C.~Meusburger and B.~J.~Schroers,
``Poisson structure and symmetry in the Chern-Simons formulation of  (2+1)-dimensional gravity,''
Class.\ Quant.\ Grav.\  {\bf 20} (2003) 2193
[arXiv:gr-qc/0301108].






\end{thebibliography}
\end{document}